\documentclass{jfm}
\usepackage{natbib}
\usepackage{graphicx}
\usepackage{amsfonts}
\usepackage{amssymb}
\usepackage{psfrag}
\usepackage{amsmath}%
\usepackage{float}
\usepackage{dcolumn}
\usepackage{bm}
\usepackage{latexsym}

\title{Influence of slip on the dynamics of two-dimensional wakes}
\author[D. Legendre, E. Lauga \& J. Magnaudet]
{\ns D\ls O\ls M\ls I\ls N\ls I\ls Q\ls U\ls E\ns L\ls E\ls G\ls E\ls N\ls D\ls R\ls E$^{1,2}$\ls,\ns E\ls R\ls I\ls C\ns L\ls A\ls U\ls G\ls A$^{3}$\\\ns \and J\ls A\ls C\ls Q\ls U\ls E\ls S\ns M\ls A\ls G\ls N\ls A\ls U\ls D\ls E\ls T$^{1,2}$}
\affiliation{
$^1$Universit\'e de Toulouse; INPT, UPS; IMFT (Institut de M\'ecanique des Fluides de Toulouse); All\'ee Camille Soula, F-31400 Toulouse, France. \,\,\,\,\,$^2$CNRS; IMFT; F-31400 Toulouse, France.  \\
$^3$Department of Mechanical and Aerospace Engineering, 
University of California San Diego,
9500 Gillman Drive, La Jolla CA 92093-0411, USA.}

\pubyear{2009}
\volume{??}
\pagerange{1--??}
\date{26 Jan 2009}
\begin{document}

\maketitle
\begin{abstract}

We study numerically the two-dimensional flow past a circular cylinder as a prototypical transitional flow, and investigate the influence of a generic slip boundary condition on the wake dynamics. We show that slip significantly delays the onset of recirculation and shedding in the wake behind the cylinder. As expected, the drag on the cylinder decreases with slip, with an increased drag sensitivity for large Reynolds numbers. We also show that past the critical shedding Reynolds number, slip decreases the vorticity intensity in the wake, as well as the lift forces on the cylinder, but increases the shedding frequency. We further provide evidence that the shedding transition can be interpreted as a critical accumulation of surface vorticity, similarly to related studies on wake instability of axisymmetric bodies. Finally, we propose that our results could be used  as a passive method to infer the effective friction properties of slipping surfaces. 

\end{abstract}

\section{Introduction}
In most continuum studies in fluid mechanics, the goal is to understand and predict the behaviour of systems with different physical phenomena at play, but there is in general very little debate regarding what the appropriate flow equations and boundary conditions are. For gases or liquids flowing under normal conditions in mesoscopic systems (millimeters or larger) bounded by solid walls, it is now universally agreed that the Navier-Stokes equations associated with the no-slip boundary condition provide an excellent description of the velocity and pressure fields in the fluid. 

Much recent experimental work has been devoted to the behaviour of fluids at sub-mesoscopic length scales, microns and nanometers (\cite{stonereview, squiresquake}). In the limit where the continuum description remains suitable, the Navier-Stokes equations still provide an appropriate predictive framework, but the validity of the no-slip boundary condition has been proven experimentally to break down in very confined systems (\cite{vinogradova99}, \cite*{granick03}, \cite{neto05}, \cite*{laugareview}). Wall slip is typically quantified by a slip length, $\lambda$, which is the fictitious distance below the slipping surface where the velocity extrapolates to zero.

Typically, surfaces in contact with a wetting fluid show very little slip, but in a non-wetting (or hydrophobic) case, the `intrinsic' slip length can reach tens of nanometers, thereby leading to significant reduction in friction on small length scales. Larger slip lengths can be obtained, for example, using  super-hydrophobic surfaces, a situation where  a high-surface energy liquid  in contact with a hydrophobic surface with significant surface roughness  spontaneously de-wets and transitions  from a state where it is everywhere in contact with the solid, to a state where it is mostly in contact with air, and is in contact with the solid only at the edge of the surface roughness (\cite*{quere05}). Measurements of laminar flow over such surfaces with roughness features in the tens of microns range show `effective' slip lengths of a few tens of microns, accompanied by noteworthy laminar drag reduction (up to 40\%, and typically 20\%, \cite*{ou04}).

Whether the boundary slip is intrinsic to the solid/surface combination, or is used as a generic but effective description of complex  surface processes  (\cite{vinogradova99,granick03}, \cite{neto05}, \cite{laugareview}), the drag reduction it implies may have a remarkable impact on flows beyond the laminar regime, and in particular on the transition to turbulence.
 This has motivated a few groups to study the effect of wall slip on flow at higher Reynolds number. A linear stability study for channel flows with slip at the wall revealed that slip delays the occurrence of flow instability, but has virtually no impact on the transient energy growth (\cite{laugacossu}). A subsequent numerical study looked at fully-developed turbulent channel flow, analyzing the influence of wall slip on turbulent structures (\cite{min04}). Surprisingly, while slip in the streamwise direction does lead to a decrease in friction, slip in the spanwise direction was found to increase turbulent drag. A related work reported computations of vortex shedding past a cylinder, both in the laminar and turbulent regimes, with either uniform slip along the cylinder or mixed slip/no-slip domains distributed along the span of the cylinder (\cite{you07}). In the laminar regime, a weak enhancement of the wake stability was observed, whereas in the turbulent regime a significant decrease of the fluctuating forces on the cylinder was noticed, as well as a narrowing of the wake. 
Finally, experiments on rolling droplets with Reynolds number in the transitional regime showed up to 15\% laminar drag reduction (\cite{gogte05}).

In this paper, we consider two-dimensional flow past a circular cylinder as a prototypical transitional flow, and ask the following questions. Generically, how does slip on the surface of the cylinder modify the transition from steady to unsteady flow? How does it change the stability diagram and the overall flow dynamics? The paper is structured as follows. We describe our numerical approach and validate our code in section 2. The results of our study, namely the influence of  slip on the dynamics of the two-dimensional wake behind a cylinder, are discussed in section 3. 
A possible application is discussed in section 4 where we propose to use our results as a passive method to infer the friction properties of slipping surfaces. We conclude with some suggestions for future work.

\section{Setup, numerical method and validation}
\label{setup}
\begin{table}
\begin{center}
\begin{tabular}{c|c|ccc}
Grid & Parameters &$C_D$&$C_L$& $St$ \\
\hline
GRID 1& $N_r \times N_\theta=80 \times 160$, $\delta=0.004 a$& 1.348&0.334&0.176\\
 GRID 2& $N_r \times N_\theta=80 \times 80$, $\delta=0.004 a$& 1.332& 0.315&0.166\\
 GRID 3& $N_r \times N_\theta=100 \times 160$, $\delta=0.002 a$& 1.347 &0.331 &0.169\\
 GRID 4& $N_r \times N_\theta=160 \times 160$, $\delta=0.001 a$& 1.339 &0.324 &0.167\\
\hline
\end{tabular}
\end{center}
\caption{Computational results obtained with three different grids in the case $Re=100, Kn=0$ (no-slip cylinder). The drag (resp. lift) coefficients are defined by dividing the time-averaged drag force (resp. the maximum of the fluctuating lift force) per unit length by $\pi\rho a^2U^2/2$. The Strouhal number is defined as $St=2a/(TU)$ where $T$ is the vortex shedding period.}
\label{tab2}
\end{table}%
We consider a two-dimensional incompressible flow past a uniformly slipping circular cylinder of radius $a$. The upstream flow is uniform, of magnitude $U$. We solve numerically the incompressible Navier-Stokes equations for the velocity, ${\bf u}$, and pressure, $p$, namely
 \begin{equation}\label{NS}
\rho\left(\frac{\partial{\bf u}}{\partial t} + {\bf u}\cdot\nabla {\bf u} \right)= -\nabla p + 2\mu \nabla \cdot {\mathbf{S}},\quad \nabla \cdot {\bf u} = 0,
\end{equation}
where $\rho$ is the fluid density, $\mu$ its dynamic viscosity and ${\mathbf{S}=(\nabla\mathbf{u} +(\nabla\mathbf{u})^T)/2}$ denotes the rate-of-strain tensor. Associated with \eqref{NS} are the impermeability and slip boundary conditions on the surface of the cylinder, which read
\begin{equation}\label{eq_1}
\mathbf{n}\cdot\mathbf{u}=0,\quad
\mathbf{n}\times\mathbf{u}=2\lambda \mathbf{n}\times(\mathbf{S}\cdot\mathbf{n}),
\end{equation}
where $\lambda$ is the cylinder slip length and $\mathbf{n}$ is the unit normal to the cylinder surface. The two dimensionless parameters characterizing the flow are the Reynolds number, $Re= 2\rho U a/\mu$, and the Knudsen number, $Kn=\lambda/a$.

The computations in this paper were carried out with the finite volume JADIM code described in detail in previous studies, e.g. \cite*{magnaudet1995} and \cite{calmet1997}.  
We use a polar grid extending up to $r_\infty=80 a$. The number of nodes is
$N_r =80$  and $N_\theta= 160$ along the radial and polar directions respectively. A 
uniform distribution in $\theta$ and a geometrical progression along  $r$ are used.
To properly capture the vorticity generated at the cylinder surface, the thickness $\delta$ of the row of cells closest to the cylinder surface is chosen as $\delta=0.004 a$, so that at least 7 cells are located within the boundary layer for the highest Reynolds number considered, the thickness $\delta_{BL}$ of the boundary layer being estimated as $\delta_{BL}\approx aRe^{-1/2}$.
The influence of these numerical parameters was carefully checked to make sure that the results are grid-independent. The example reported in Table \ref{tab2} shows the drag and lift coefficients and the Strouhal number obtained for $Re=100$ in the no-slip case with the grid used in this study, called GRID1. These results are compared with those given by three other grids (see also Table \ref{tab1}). It may be observed that the changes in $N_r$, $N_\theta$ as well as in $\delta$ do not induce significant modifications of the drag, lift and shedding frequency (most of the variation on the last two quantities are due to the marginal sampling time rather than to the grid). Similar results were obtained for the surface vorticity near the rear stagnation point. This region is very sensitive to the grid characteristics because the sign of the vorticity changes due to separation, and a very good agreement was also observed between the predictions provided by the four grids. The code has been extensively validated in the past, especially for bubbles and solid spheres (e.g. \cite{magnaudet1995}). In order to provide an extra validation in the present configuration, the values of the drag coefficient, the maximum value of the lift coefficient and the Strouhal number are reported for $Kn=0$ (no-slip) in Table \ref{tab1} and compared with a selection of previous results available in the literature.

Let us remind that the flow past a no-slip circular cylinder first separates for a critical Reynolds number $Re_1$ slightly less than 7 (\cite*{chen95}), while a second transition where the flow becomes time-dependent and vortices of alternating signs are shed from either side of the cylinder occurs at another critical Reynolds number $Re=Re_2$ which is between 46 and 50 (\cite{williamson96}), the Strouhal number $St(Re_2)$ being about $0.13$. Present computations indicate $Re_1=6.5$, $Re_2=47.5$ and $St(Re_2)\approx0.131$, in excellent agreement with previous studies.
\begin{table}
\hspace{-1.2 cm}\begin{tabular}{c|c|ccccccc}
$Kn=0$& &$Re=5$&20& 50&100&200&500&800 \\
\hline
&  DNS1 & 4.116&2.045&-&-&-&-&-\\
$C_D$ & \quad\quad  2D (resp. 3D) DNS2\quad \quad &-&-& -& \quad 1.253 (1.240)& \quad 1.321 (1.306)&-&-\\
& This study &4.065&2.035&1.445& 1.350 & 1.345 &1.379&1.391 \quad\\ 
\hline
$C_L$ & 2D (resp. 3D) DNS2 &-&-&-&0.39 (0.36)&0.76 (0.64)&-&- \\
& This study &0&0&0.066&0.334&0.70&1.11&1.14\\
\hline
 & \quad  Experiments 
 & -& -& 0.123&0.164&0.197&-&-\\
$St$ & 2D (resp. 3D) DNS2& -&-&-&0.165 (0.164)&0.198 (0.181)&-&-\\
&  DNS3 & & &0.140&0.179&0.206 
&-&-\\
& This study &0&0 &0.133&0.176&0.204&0.222&0.224\\
\hline
$Kn=\infty$& & $Re=5$& 20 & 50&100&200&500&800 \\
\hline
$C_D$ &This study&3.15&1.33 & 0.712 & 0.415 &0.228&0.0973&0.0618\\
 \hline

\end{tabular}
\caption{Drag coefficient, lift coefficient and Strouhal number as a function of the Reynolds number in the case of a no-slip cylinder ($Kn=0$) and a shear-free cylinder ($Kn=\infty$). Results for the no-slip cylinder are compared with previous results from DNS (DNS1, DNS2, and DNS3 are from \cite{dennis70}, \cite{persillon98} and \cite{karniadakis89}, respectively) and experiments (\cite{williamson88}).}
\label{tab1}
\end{table}

Numerically, the slip condition (\ref{eq_1})
is implemented in the following manner. In a system of polar coordinates, the tangential strain rate at the surface is $S_{r \theta}{\mid}_{r=a}= (\partial u_\theta/\partial r- u_\theta/a){\mid}_{r=a}$. Evaluating the Taylor expansion of the tangential velocity in the vicinity of $r=a$ at the centre of the first two rows of cells surrounding the cylinder ({\it i.e.} at distances $d_1$ and $d_2$ ($d_1<d_2$) from the surface), and equating these expansions with the numerical values found at the previous iteration for $u_{\theta 1}=u_{\theta}{\mid}_{r=a+d_1}$ and $ u_{\theta 2}=u_{\theta}{\mid}_{r=a+d_2}$ respectively, yields a second-order accurate explicit expression for $\partial u_\theta/\partial r{\mid}_{r=a}$.
Combining the resulting approximation with the second equation in  (\ref{eq_1}), the surface velocity $u_{\theta}{\mid}_{r=a}$ is obtained as
\begin{eqnarray}\label{eq_3}
 u_{\theta}{\mid}_{r=a} = \left( \frac{d_2}{d_1 (d_2- d_1)} u_{\theta 1} - \frac{d_1}{d_2 (d_2- d_1)} u_{\theta 2} \right)/
 \left(\frac{1}{\lambda} +\frac{1}{a} +\frac{d_1 +d_2}{d_1 d_2}\right)
\end{eqnarray}
Using (\ref{eq_3}), the surface shear stress $2 \mu S_{r \theta}{\mid}_{r=a}$ involved in the local momentum balance is finally obtained through the above approximation for $\partial u_\theta/\partial r{\mid}_{r=a}$.
\section{Influence of slip on the wake dynamics}
\label{results}

\subsection{Stability diagram}
\begin{figure}
\begin{center}
\includegraphics[width=350 pt]{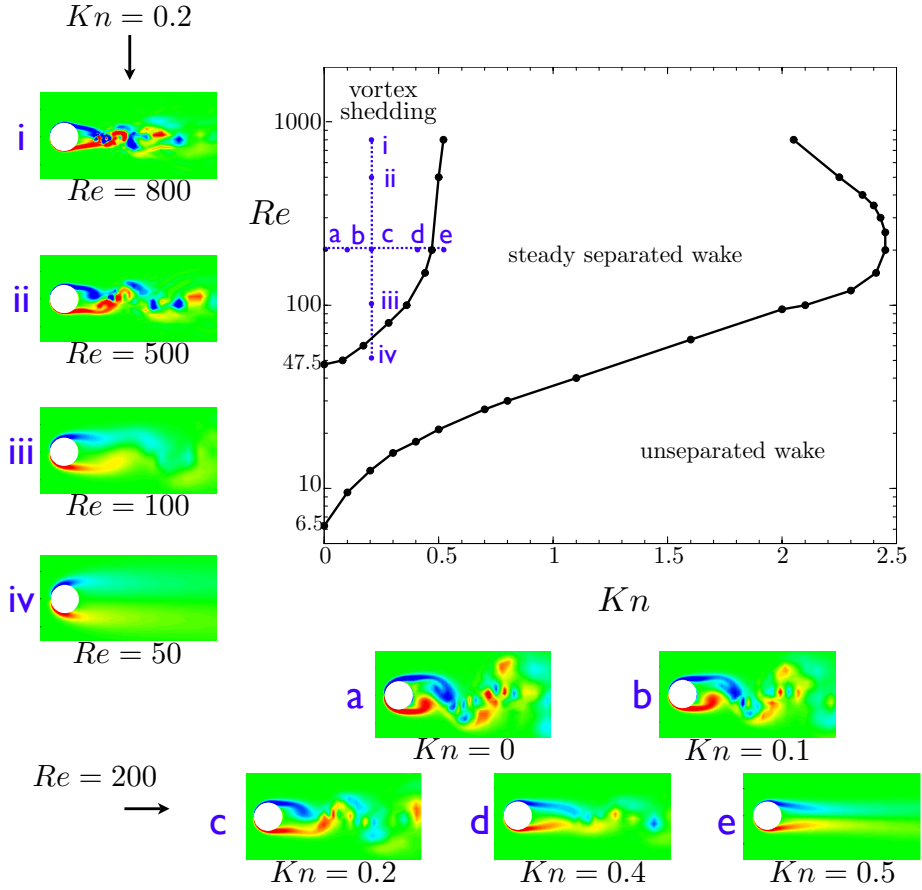}
\end{center}
 \caption{Stability diagram in the $(Re,Kn)$ space and structure of the wake vorticity for some selected values of the Reynolds number, $Re$, and Knudsen number, $Kn$ (color scheme is the same for all figures; red is positive vorticity). For each $Kn$ (resp. $Re$), the transition curves were obtained by varying $Re$ (resp. $Kn$) until the results for two consecutive values were found to lie on each side of the corresponding transition; a linear interpolation was then used to determine more accurately the corresponding critical $Re$ or $Kn$. The closed circles on the transition curves indicate the pairs ($Re, Kn$) at which the flow was explored for nearly-critical conditions. Separation and vortex shedding were detected by examining the occurrence of a change of sign in the surface velocity and that of a nonzero lift force, respectively.}
\label{mainfig}
\end{figure}

The main result of our investigation is illustrated in figure \ref{mainfig}, where we display the stability diagram for the wake in the $(Re,Kn)$ plane. The regions separated by the two solid lines are the subdomains where the wake is, in the order of increasing Reynolds number, steady and unseparated, steady and separated, and unsteady ({\it i.e.} in the vortex shedding regime). The main conclusion inferred from figure \ref{mainfig} is that slip delays both the onset of separation and the onset of vortex shedding. The effect is significant, and small values of $Kn$ lead to large changes in the critical Reynolds number $Re_2$, with a very sharp increase for $Kn\gtrsim 0.5$. Note also that the critical line for the onset of separation behind the cylinder is not monotonic. That is, for Knudsen numbers typically in the range $[2., 2.5]$, the flow separates only within a finite range of Reynolds number, say $[Re_{1l},Re_{1u}]$, whereas it remains unseparated both for $Re<Re_{1l}$ and for $Re>Re_{1u}$ (see the discussion in 3.3).
We also illustrate in figure \ref{mainfig}, as insets, the flow in the wake by displaying vorticity snapshots for some selected values of $Re$ and $Kn$. For a given $Re$, increasing $Kn$ leads to a decrease in the vorticity intensity in the boundary layer and wake, and a narrowing of the wake due to the decrease of the separation angle. In the case $Re=200$, the separation angle varies from $72.6^\circ$ for $Kn=0$ to $27.2^\circ$ for $Kn=0.5$ and finally falls to zero for $Kn=2.45$.

\subsection{Global flow characteristics}
\label{sec:draglift}
\begin{figure}
\begin{center}
{\includegraphics[clip,width=350 pt]{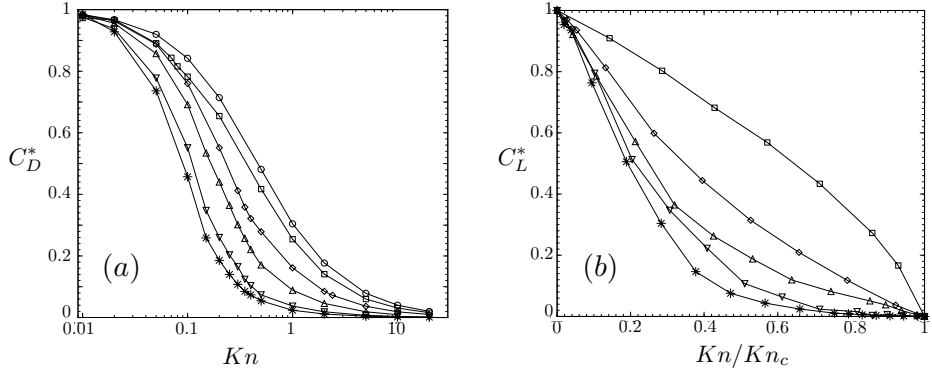}}
\end{center}
\caption{Drag and lift coefficients as a function of the Knudsen number for $Re=20$ ($\circ$, only in (a)), 50 ($\square$), 100 ($\diamond$), 200 ($\triangle$), 500 ($\triangledown$) and 800 ($\ast$). (a): normalized drag coefficient, $C_D^{*}(Kn)=(C_D(Kn)-C_D(\infty))/(C_D(0) -C_D(\infty))$; (b): normalized lift coefficient, $C_L^{*}(Kn)=C_L(Kn)/C_L(0)$. In (b), the Knudsen number is normalized by its critical value, $Kn_c(Re)$, above which the flow is no longer unsteady ($Kn_c$=0.09, 0.38, 0.47, 0.49 and 0.53 for $Re$=50, 100, 200, 500 and 800, respectively).}
\label{draglift}
\end{figure}

The changes in the forces on the cylinder as a function of wall slip are shown in figure \ref{draglift}a-b. In figure \ref{draglift}a, we display the normalized time-averaged drag coefficient $C_D^{*}(Kn)=(C_D(Kn)-C_D(Kn=\infty))/(C_D(Kn=0) -C_D(\infty))$. The values corresponding to the no-slip limit ($Kn=0$) and to perfect slip ($Kn=\infty$) are available in Table 2. As expected, the drag on the cylinder decreases monotonically with increasing wall slip, with a steeper decrease as the Reynolds number goes up. Typically, the normalized drag coefficient crosses the median value 0.5 for $Kn\approx 0.5$ when $Re=20$ and for $Kn\approx 0.1$ for $Re=800$. In other words, the wake becomes more sensitive to a small amount of slip as inertial effects increase. We also see on Table 2 that the change in the drag coefficient between the no-slip and infinite slip cases is small for small Reynolds numbers, but becomes dramatic as the Reynolds number increases. 
For Reynolds numbers beyond $Re=50$, the amount of drag reduction is larger than 50\%, and is above 80\% for $Re > 200$. 
In figure \ref{draglift}b, we display the normalized lift coefficient $C_L^{*}(Kn)=C_L(Kn)/C_L(Kn=0)$, as a function of the Knudsen number normalized by the critical value $Kn_c$ beyond which the wake no longer sheds vortices (and therefore displays no lift). Again, we observe a smooth decay of the lift force with increasing wall slip, and an increase of the sensitivity of the wake as the Reynolds number is increased. The origin of the drag reduction with increasing slip is  twofold. First, for a sufficiently large Reynolds number, the vorticity at the surface decreases from $O(Re^{1/2})U/a$ for $Kn=0$ to $O(1)U/a$ for $Kn=\infty$ (see figure \ref{vorticity}), and the skin friction decreases accordingly. Second, this decrease in the surface vorticity reduces or even suppresses separation at the cylinder surface. Therefore the fore-aft asymmetry of the surface pression distribution decreases, thereby reducing the form drag. As less vorticity is produced at the cylinder surface when $Kn$ increases, the strength of the vortices shedded in the wake decreases with increasing slip, and so does the lift force.

\begin{figure}
\begin{center}
 \includegraphics[width=190 pt]{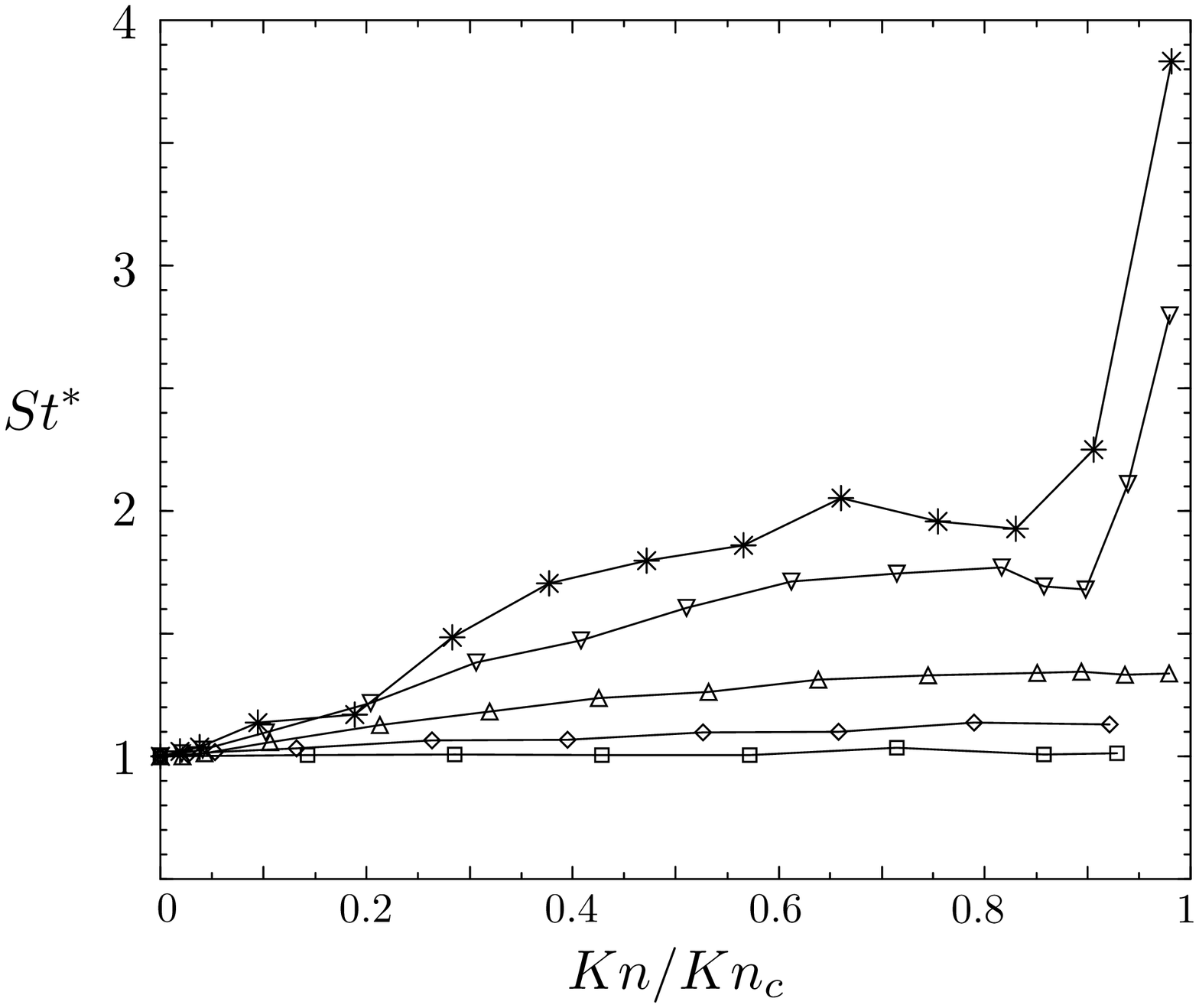}
\end{center}
 \caption{Normalized Strouhal number $St^{*}(Kn)=St(Kn)/St(0)$, as a function of $Kn$ for $Re=50$, 100, 200, 500 and 800. $St$ was computed by considering the most energetic peak in the lift force spectrum. The definition of symbols and that of $Kn_c$ are similar to those in figure \ref{draglift}.
 }
\label{strouhal}
\end{figure}

Further insight on the wake  dynamics  can be gained by computing the shedding dominant frequency $1/T$ as a function of slip. The results are illustrated in figure \ref{strouhal} where we display the normalized Strouhal number for the wake, $St^{*}(Kn)=St(Kn)/St(0)$, as a function of $Kn/Kn_c$ for different values of $Re$. The shedding frequency increases with $Kn$, with up to a factor of 4 increase at large Reynolds number. Hence, although slip leads to a decrease of the force amplitudes on the cylinder, it increases the frequency at which these force components oscillate. This is because the time rate-of-change of the vorticity at the cylinder surface (which is proportional to $St$) results from both the diffusion of vorticity across the boundary layer and its advection along the cylinder surface. While the former is almost unaffected by slip, the strength of the latter increases linearly with the tangential velocity at the cylinder surface. Combining (\ref{eq_1}) with an estimate of the boundary layer thickness $\delta_{BL}/a=\alpha Re^{-1/2}$ with $\alpha =O(1)$, one can show that the magnitude of the tangential velocity scales as $\alpha Re^{1/2}/(\alpha Re^{1/2}+1+Kn^{-1})$, so that $St$ increases with $Kn$ when $Re$ is kept fixed.

\subsection{Surface vorticity}
\label{sec:vorticity}

There are strong indications that the formation of a standing eddy and the occurrence of wake instability past axisymmetric bodies arises when the vorticity accumulated around the body, {\it i.e.} the difference between the amount of vorticity generated at the body surface and that which is evacuated downstream in the wake, exceeds some critical value (\cite*{leal1989}). This idea  explains why the wake past rising spherical gas bubbles remains axisymmetric for all Reynolds numbers, while the axial symmetry breaks down when the bubble deformation exceeds some critical value, vorticity at a perfect-slip surface being directly proportional to the surface curvature (\cite{magnaudet2007}). In the latter paper it was shown that, once expressed in terms of the maximum vorticity at the body surface, $\omega_{max}$, the criterion predicting whether the wake past a solid sphere is stable or not is identical to that predicting wake instability for a shear-free bubble. This criterion indicates that the wake is unstable when $\omega_{max}(Re)/(U/a)>g(Re)$, where $g$ is a function that weakly depends on the Reynolds number. In other terms, the strength of the vorticity at the body surface is the key quantity that determines the stability of the wake, irrespective of the specific dynamic boundary condition at the surface.

\begin{figure}
\begin{center}
\includegraphics[width=260 pt]{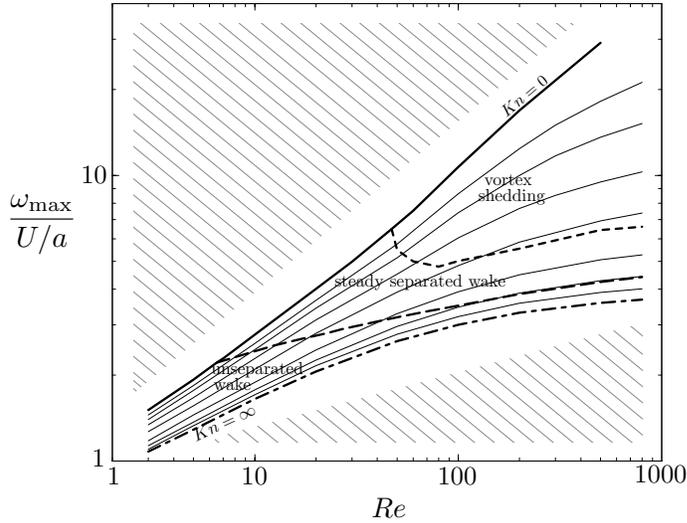}
\end{center}
\caption{Maximum dimensionless vorticity $\omega_{max}/(U/a)$ at the cylinder surface as a function of the Reynolds number (log-log scale). The thick solid and dash-dotted lines correspond to $Kn=0$ and $Kn=\infty$, respectively; the thin solid lines correspond to $Kn=$ 0.05, 0.1, 0.2, 0.4, 1, 2.2 and 5, respectively. The long and short dashed lines delimit the regions where the wake is unseparated, steady and separated, and unsteady, respectively. These lines were obtained using the procedure described in the caption of figure 1.}
\label{vorticity}
\end{figure}

 It is of interest to examine how the same argument works in the present configuration. For this purpose, we show in figure \ref{vorticity} the quantity $\omega_{max} a/ U$, for the two transitions displayed in figure \ref{mainfig}, as well as along iso-$Kn$ curves, as a function of the Reynolds number. Note that the curve corresponding to $Kn=0$ exhibits a $1/2$ slope for large enough Reynolds number, reflecting the fact that the normalized vorticity at a no-slip surface grows like $Re^{1/2}$. The results shown in figure \ref{vorticity} reveal that along the shedding transition curve, the Reynolds number increases by a factor of 20, whereas the critical vorticity, although dependent on $Re$, displays little variation and remains in the range 
 $5 \lesssim  \omega_{max} a/ U \lesssim 7$. Hence, present results offer additional evidence that the shedding transition occurs when the surface vorticity exceeds some critical value that weakly varies with the Reynolds number, irrespective of the dynamic boundary condition at the cylinder surface. In particular one can infer from figure \ref{vorticity} that the wake past a shear-free cylinder (for which $\omega_{max} a/ U=4$, as shown by assuming a potential flow solution with a vortex sheet at the cylinder surface) remains steady and unseparated for arbitrarily large  Reynolds numbers. The same argument allows us to understand why, according to figure \ref{mainfig}, the wake separates only within a finite range of Reynolds number, $[Re_{1l},Re_{1u}]$, when $Kn$ exceeds some critical value: vorticity generation at the surface increases only weakly with $Re$ in presence of significant slip, while the vorticity flux advected downstream in the wake increases linearly with the upstream velocity (and therefore with $Re$), implying the existence of an upper value of $Re$ beyond which there is not enough vorticity accumulated behind the cylinder for the wake to separate (\cite{leal1989}).

\section{Application to the effective friction of slipping surfaces}
\label{method}

The results presented in figure \ref{mainfig} could be used as a method to passively infer the friction properties of surfaces prone to slip. For example,  given the potentially large drag reduction resulting from super-hydrophobic coatings, there is growing interest in obtaining quantitative information regarding their overall friction properties, and the effectiveness of particular geometrical designs (length scales and morphology). The methods used so far in the literature involve real friction measurements, where a particular surface is modified to become super-hydrophobic, a pressure gradient is applied, and the change in the flow rate is measured (\cite{ou04}). These are arguably difficult experiments and it would be useful that simpler methods be available.

In this spirit, we propose to coat a cylinder with the studied slipping surface, and to place it in a uniform flow field with known upstream velocity $U$. As the flow speed is increased beyond a critical value, the flow will become unsteady, and we can then exploit our numerical results  to infer the effective slip length on the surface, without performing any actual friction measurement. With the knowledge of the critical value of $U$, and hence $Re$, at which vortex shedding sets in, one reads the corresponding value of $Kn$ from figure \ref{mainfig}, and  infers the effective slip length describing the surface.  According to the results displayed in figure \ref{mainfig}, this will work as long as $Kn\lesssim 0.5$.

For  super-hydrophobic surfaces, this  method will be effective provided two hypotheses are satisfied. We first need to ensure that the surface remains super-hydrophobic, and does not transition back to a wetted state. We further need to ensure that the effect be sufficiently strong to be experimentally measurable.
Regarding the first point, it is necessary that the flow-induced pressures, which are on the order of $ \rho U^2$, are small enough to prevent transition to the wetting state. For typical super-hydrophobic surfaces made of a dense population of pillars with size $r$, the critical pressure is a capillary pressure, on the order of $\sigma /r$, where $\sigma$ is the surface tension of the liquid in air (\cite{bartolo06}). Wetting will therefore be avoided if $\rho U^2 \lesssim \sigma /r$, or $Re \lesssim a/\sqrt{\ell r}$,
where $\ell = \mu^2/\rho\sigma$ is the Ohnesorge length for the liquid. 
Regarding the second point, we need the slip length  to be sufficiently large for the effect to be detectable. According to previous work by \cite{ybert07}, we expect $\lambda \sim d$, where $d$ is the distance between the pillars composing the super-hydrophobic surface. To fix ideas, if we wish the transition to occur at a particular $Kn$, we need $a\approx d/Kn$. This limit will be compatible with the constraint found above if $Re \, Kn \lesssim d/\sqrt{\ell r}$
and, with $r\sim d$, we need therefore to be in the limit where $(Re \, Kn)^2 \lesssim {r}/{{\ell}}$.
For an air-water system, we have $\ell\approx 14$ nm. The critical Reynolds number for, say, $Kn=0.1$ is about $60$, which is sufficiently above $Re_2(Kn=0)=47.5$ to be measured without ambiguity, so $r$ and $d$ need to be at least a few microns, and the proposed experiment can be achieved in a microfluidics device. 
\section{Concluding remarks}
\label{end}

In this work, we performed direct numerical simulations to study the influence of generic slip boundary conditions on the dynamics of the two-dimensional wake behind a circular cylinder. Specifically, we have shown that slip: 
 (1) delays the onset of recirculation and vortex shedding in the wake;
(2) decreases both drag and lift;
(3) decreases the vorticity intensity in the near wake; 
(4) increases the shedding frequency. 
We have also discussed a practical application of our results and proposed that they could be used as a passive method to infer the effective friction properties of slipping surfaces. 

The most severe limitation of the current work is the two-dimensional  assumption for all Reynolds numbers. In the case of a long, no-slip cylinder, three-dimensional effects are known to become important when the Reynolds number reaches $Re\approx 190$ \cite{williamson96}. However, since these three-dimensional effects occur well into the unsteady regime, the wake is likely to remain two-dimensional close to the shedding transition, and our critical curves $Re_1=f(Kn)$ and $Re_2=g(Kn)$ should not be affected by these effects.  Regarding the applicability of our results as a method to passively estimate wall slip, other effects such as geometrical confinement  would lead to changes in the details of the results shown in figure \ref{mainfig}, and each experimental setup would be characterized by its own transition curve to be obtained numerically, but the general principle remains valid. Moreover, in general, most complex surfaces cannot be adequately modeled by a single homogeneous slip coefficients, but instead a detailed look at the geometrical and physico-chemical nature of the interface is required. Results  in the case of micro-textured surface  will be reported in future work.\\

This work was funded in part by the US National Science Foundation (grants CTS-0624830 and CBET-0746285 to Eric Lauga).

\bibliographystyle{jfm}
\bibliography{slip}
\end{document}